\documentclass[aps,pre,floats,floatfix,twocolumn,superscriptaddress]{revtex4}

\usepackage{latexsym,amsmath}
\usepackage{amsbsy}
\usepackage[pdftex]{graphicx}

\usepackage{amssymb}
\usepackage{epstopdf}
\usepackage{placeins}
\usepackage[usenames]{color}

\begin{document}

\title{Regulation of burstiness by network-driven activation}

\author{Guillermo Garc{\'i}a-P{\'e}rez}
\author{Mari{\'a}n Bogu{\~n}{\'a}}
\author{M. \'Angeles Serrano}
\affiliation{Departament de F{\'\i}sica Fonamental, Universitat de Barcelona, Mart\'{\i} i Franqu\`es 1, 08028 Barcelona, Spain}

\date{\today}

\begin{abstract}
We prove that complex networks of interactions have the capacity to regulate and buffer unpredictable fluctuations in production events. We show that non-bursty network-driven activation dynamics can effectively regulate the level of burstiness in the production of nodes, which can be enhanced or reduced. Burstiness can be induced even when the endogenous inter-event time distribution of nodes' production is non-bursty. We found that hubs tend to be less controllable than low degree nodes, which are more susceptible to the networked regulatory effects. Our results have important implications for the analysis and engineering of bursty activity in a range of systems, from telecommunication networks to transcription and translation of genes into proteins in cells.
\end{abstract}

\maketitle
While human perception tends to appreciate regularity in the happening of events, evidence reveals that in many systems events cluster in bursts, {\it i.e.}, accumulations of a large number of rapidly occurring events during short time intervals separated by silent periods. Bursts have been experimentally and empirically observed in profusion and many models have been proposed to explain and generate them. For example, traces of human activity are to a great extent imprinted with burstiness. In particular, bursts and heavy-tails pervade the diversity of human communication channels, from written text~\cite{Church:1995,Serrano:2009a,Altmann:2009}, to the frequency of letter~\cite{Malmgren:2009} and e-mail correspondence~\cite{Barabasi:2005,Malmgren:2008}, and cell phone calls~\cite{Karsai:2012qf,Jiang:2013}.

In a different context, key processes in single-cell biology, such as transcription and translation, are particularly prone to progress in bursts. For instance, mRNA synthesis proceeds in short but intense outbreaks beginning when coding genes transition from an inactive to an active state~\cite{Raj:2006}. These bursts have been suggested to be able to affect the expression of essential genes and even the switching and rhythms of cellular states~\cite{Schultz:2007,Zong:2010,Toner:2013}. This may raise questions about how cells are able to maintain function in the face of such unpredictable fluctuations explained by different single gene burst generation mechanisms~\cite{Thattai:2001,Paulsson:2004,Dobrzynski:2009}, with the hypothesis that those bursts could be buffered by local or bulk degradation mechanisms~\cite{Raj:2006}. It is interesting to explore whether large-scale system mechanisms, like the interconnection of genes in regulatory networks of interactions, have the capacity to produce similar buffering effects.

In this work, we show indeed that endogenous burstiness in the production of nodes can be both enhanced or reduced by stochastic network-driven non-bursty interactions, which activate/inactivate nodes dynamically. We observe a strong anti-correlation of these regulation phenomena with node degree, such that the burstiness of lower degree nodes is always more susceptible to being raised by dynamic network influence, while hubs tend to experience a decreased burstiness in their endogenous production, which helps to maintain a stable bulk level. Conversely, the production sequence of individual nodes of all degrees can present bursts even when their endogenous production pattern follows a non-bursty inter-event time probability distribution, and we observe again a strong anti-correlation of the induced burstiness with node degree. 

Note that here we are reversing the topical question about the macroscopic collective effects of burstiness on dynamical processes running on complex networks. The bursty activity of nodes, both in models of opinion formation \cite{Fernandez-Gracia:2011,Takaguchi:2011} and epidemic spreading \cite{Vazquez:2007,Karsai:2011,Min:2011}, has been observed to induce a slowing down of the progression of the dynamical process. Here, we focus instead on the regulatory effects of networked dynamical interactions on the endogenous bursty activity of individual nodes.

We consider nodes interconnected forming a complex network that can present two states, active and inactive. During the active state, each node produces events according to a certain endogenous inter-event time probability distribution function $\psi(t)$. The activation dynamics of nodes is governed by a propagating process in the network. We chose an Inactive-Active-Inactive (IAI) dynamical process akin to the standard Susceptible-Infected-Susceptible epidemic spreading model~\cite{Anderson:1991}, so that active nodes inactivate at rate $\lambda_i$ and inactive ones become active upon contact with active ones at rate $\lambda_a$ per active contact. In all cases, these processes are assumed to follow Poisson statistics. An endemic state of active nodes can be sustained whenever the ratio $\lambda_a/\lambda_i$ is above a critical value that, in general, depends on network topology~\cite{Romualdo:2001,Boguna:2002}. Once at steady state, the effective activation rate of a node depends on the topological configuration of the network, with a temporal average value on uncorrelated networks given by~\cite{Romualdo:2001,Boguna:2002}
\begin{equation}\label{eq:lambda_k}
\lambda_{eff}(k) = k \lambda_{a} \bar{\rho}=k \lambda_{a}\left< \sum \limits_{k'} \rho_{k'}(t) \frac{ k' P(k')}{\left< k \right>} \right>.
\end{equation}
Here $k$ is the node degree and $\bar{\rho}$ is a temporal average of a degree-weighted prevalence depending on $P(k)$ --the degree distribution of nodes in the network--, the average degree $\left< k \right>$, and the fraction of nodes with degree $k'$ in the active state $ \rho_{k'}(t)$. This temporal average fluctuates around a constant value in the stationary state of the endemic phase (see Appendix~\ref{sec:dynamics}), so that $\lambda_{eff}(k)$ is proportional to degree. 

This result allows us to focus the problem on the study of a single node that changes state following Poisson processes with rates $\lambda_{eff}(k)$ for activation and $\lambda_{i}$ for inactivation and that, when active, produces events according to a general distribution $\psi(t)$ with average $\langle t_p \rangle$. For such a node, and any $\psi(t)$, we calculated analytically the effective inter-event time probability density function of production events ${\phi}(t,k)$ (see Appendix~\ref{sec:phi}). Here, we report the result in the Laplace space $\hat{\phi}(s,k) \equiv \int_0^{\infty} e^{-st}\phi(t,k)dt$,
\begin{equation}\label{eq:phi}
\hat{\phi}(s,k) = \hat{\psi}(s+\lambda_{i}) + \frac{\frac{\left[ 1 - \hat{\psi} (s + \lambda_{i}) \right]^{2}}{\left< t_p \right> \left( s + \lambda_{i} \right)^{2}} }{\frac{s + \lambda_{eff}(k)}{\lambda_{eff}(k) \lambda_{i}} - \frac{\left[ 1 - \frac{1 - \hat{\psi} (s + \lambda_{i})}{\left< t_p \right> \left( s + \lambda_{i} \right)} \right]}{s + \lambda_{i}}}
\end{equation}
where we have used that the probability density for the interval between the activation time and the first production event is given by $\psi^{0}(t) = \Psi(t)/\left< t_{p} \right>$, where $\Psi(t)=\int_t^{\infty} \psi(t')dt'$ is the survival function. The rationale for this choice is that the activation event can be regarded as taking place between two consecutive production events. Since we do not have any other information, we assume that the time interval between those two events is greater than the time elapsed between the activation event and the first production. As a consequence, its probability density must be proportional to $\Psi(t)$, and the denominator simply normalises the distribution~\cite{cox:1970}. Hence, effective inter-event times of production events are ruled by the interplay of three key factors: the endogenous production statistics $\psi(t)$, the first production event statistics $\psi^{0}(t)$, and the activation dynamics through the effective activation rate that depends linearly on a node's degree $\lambda_{eff}(k)$. 

To measure the burstiness of production, we use the burstiness coefficient $B=(CV_\tau-1)/(CV_\tau+1)$ as defined in~\cite{Goh:2008}, where $CV_\tau$ is the coefficient of variation $CV_\tau=\sigma_\tau/\mu_\tau$ and $\mu_\tau$ and $\sigma_\tau$ are the mean and standard deviation of the time between consecutive production events. With this definition, $B=1$ corresponds to a strongly bursty production, $B=0$ to a neutral one following a Poisson statistics, and $B=-1$ to a periodic signal. As an example, we now particularize to the case of a Weibull distribution for production events
\begin{equation}\label{eq:weibull}
\psi(t) = \sqrt{\frac{\beta}{2 t}} e^{- \sqrt{2 \beta t}}
\end{equation}
with scale parameter $(2\beta)^{-1}$ controlling the spread of the distribution, and shape parameter $1/2$ implying a heterogeneous non-Poisonian distribution such that all nodes have an endogenous production of events that clusterize in time. The endogenous burstiness for a node continuously producing according to Eq.~\eqref{eq:weibull} is $B= (\sqrt{5}-1)/(\sqrt{5}+1)=0.382$, independent of $\beta$.

The effective value of the burstiness of a node with interrupted production due to the network-driven activation/inactivation dynamics depends on its degree, $B(k)$. To compute it, we need to calculate the degree-dependent average and standard deviation of production inter-event times, which can be easily evaluated through simple derivatives of Eq.~(\ref{eq:phi}) evaluated at $s=0$ (see Appendix~\ref{sec:phi_weibull}). The degree dependent coefficient of variation reads

\begin{figure}[ht]
\includegraphics[width=\linewidth]{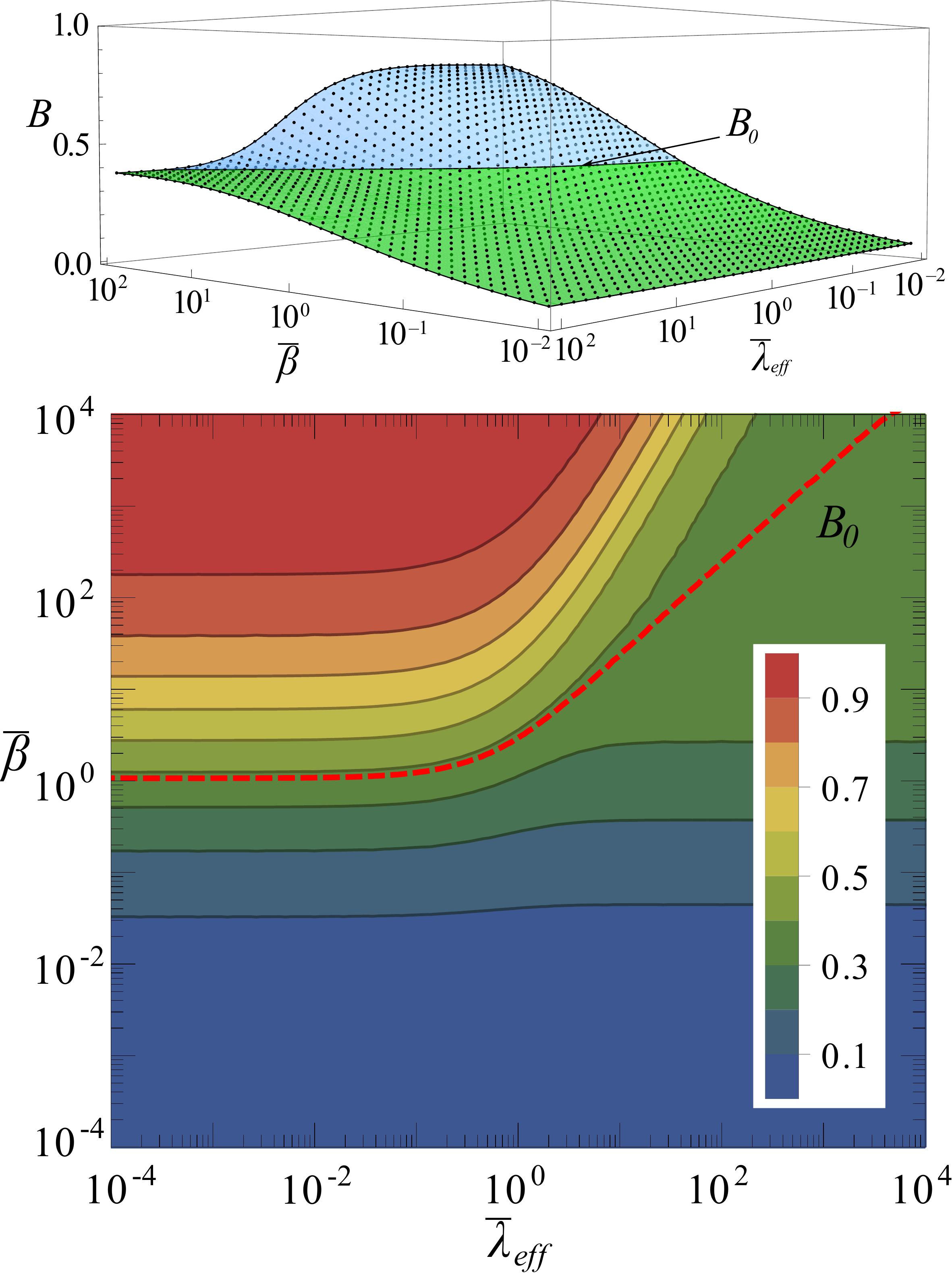}
\caption{\label{fig:theo_reduction1} {\bf Regulation of burstiness.} {\bf a}. Comparison between simulation data (dots) and the analytical calculation (solid surface and line) of the effective burstiness $B(\bar{\beta},\bar{\lambda}_{eff})$, $B>B_{0}=0.382$ in the blue region and $B<B_{0}=0.382$ in the green one. {\bf b}. Projection of burstiness surface in {\bf a} in the $\bar{\beta},\bar{\lambda}_{eff}$ plane. Colors in the contour plot code for different values of $B$.}
\end{figure}

{\scriptsize 
\begin{equation}
CV_\tau(k) = \ \sqrt{\frac{2}{1 - \sqrt{\pi \frac{\bar{\beta}}{2}}e^{\frac{\bar{\beta}}{2}}\text{erfc}\left( \sqrt{\frac{\bar{\beta}}{2}} \right)} - 1 + 2 \bar{\beta} \left( \frac{1}{\left( 1 + \bar{\lambda}_{eff}(k) \right)^{2}} - 1 \right)},
\label{bkbr}
\end{equation}}
where we have redefined $\bar{\beta}=\beta/\lambda_i$ and $\bar{\lambda}_{eff}(k)=\lambda_{eff}(k)/\lambda_i=k\bar{\lambda}_{a}\bar{\rho}$. As a validation, we display in Fig.~\ref{fig:theo_reduction1}a the analytical result for $B$ based on Eq.~\eqref{bkbr} along with the simulation of a single node whose state changes following Poisson processes with rates $\bar{\lambda}_{eff}$ for activation and $\lambda_i = 1$ for inactivation, and that intrinsically produces events according to Eq.~\eqref{eq:weibull} with rate $\bar{\beta}$. The agreement between the analytical surface and the simulation points is excellent. Both results prove that the level of endogenous burstiness for continuous production $B_0=0.382$ can be both increased and decreased, as shown also in the projection of the analytical surface on the $\bar{\beta}-\bar{\lambda}_{eff}$ plane in Fig.~\ref{fig:theo_reduction1}b. The effective burstiness ranges always between zero burstiness (exponential inter-event time distribution) in the limit $\bar{\beta} \rightarrow 0$ and the maximum of $1$ when $\bar{\beta} \rightarrow \infty$. For high values of $\bar{\lambda}_{eff}$, $B$ is dominated by the interplay of the endogenous production statistics $\psi(t)$ and the first production event statistics $\psi^{0}(t)$. The range of $\bar{\beta}$ values associated to effective burstiness around $B_0$ widens with increasing $\bar{\lambda}_{eff}$ and the endogenous production level is recovered only when both $\bar{\beta}$ and $\bar{\lambda}_{eff}$ are increased simultaneously. In contrast, the effective burstiness raises for low values of $\bar{\lambda}_{eff}$ due to the effect of increased inactivation periods, such that for each $\bar{\lambda}_{eff}$ the level of effective burstiness increases with $\bar{\beta}$. These results are qualitatively the same for any distribution $\psi(t)$ (see Appendix~\ref{sec:limits} for a detailed analysis).

In line with these arguments, and from the proportional dependence of $\lambda_{eff}(k)$ with degree Eq.~\eqref{eq:lambda_k}, one can expect that the effective burstiness of higher degree nodes is reduced while lower degree nodes can be modulated. To confirm this, we measured $B(k)$ from simulations on a network with $N=10^4$ nodes (simulation details in Appendix~\ref{sec:simulation}). Although our results are valid in the variety of uncorrelated network topologies, we used a scale-free network with characteristic exponent $\gamma=2.5$. 

\begin{figure}[t]
\includegraphics[width=\linewidth]{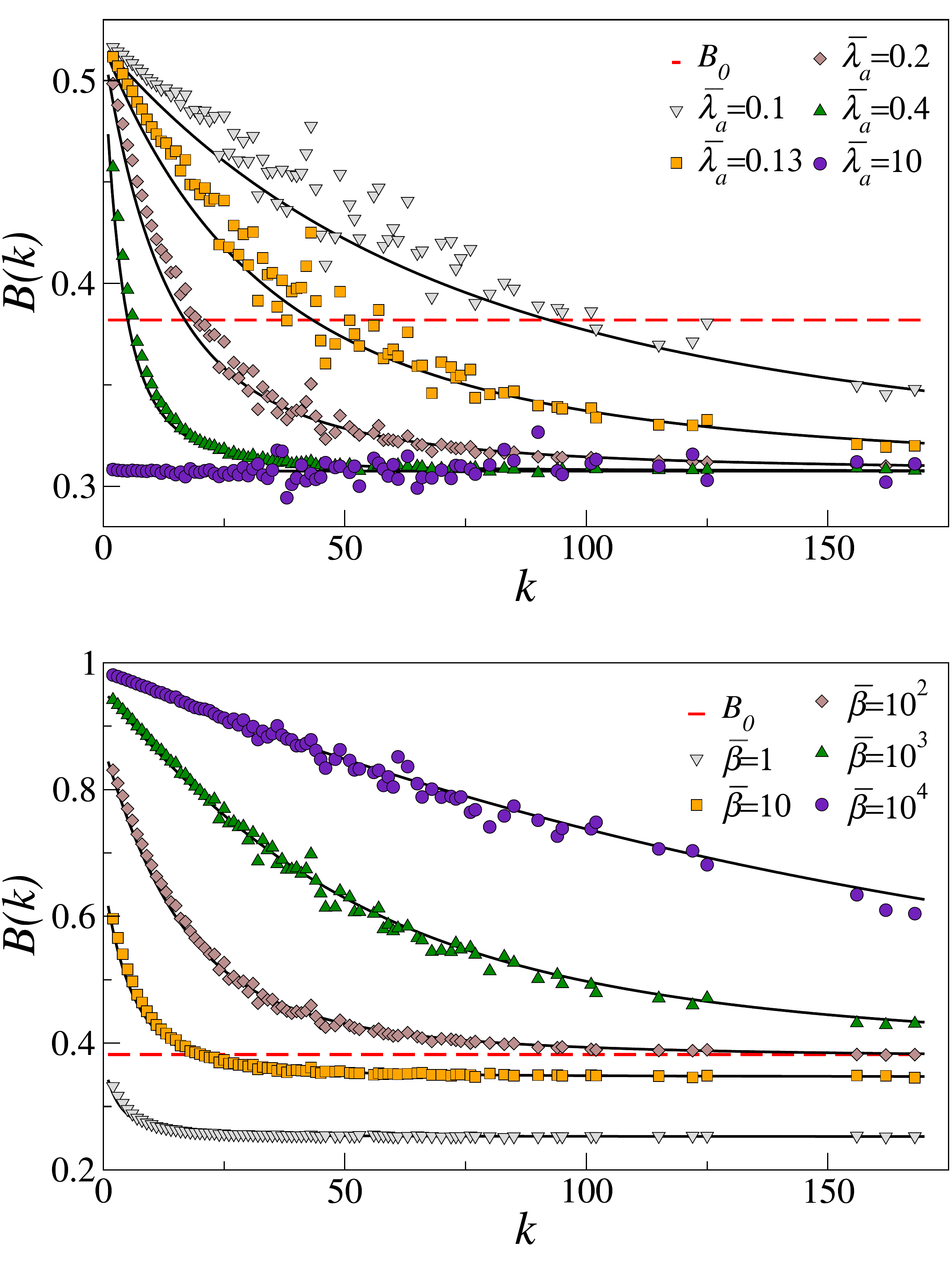}
\caption{\label{fig:theo_reduction2} {\bf Network-driven regulation of burstiness.} {\bf a}. $B(k)$ for different values of the activation rate $\bar{\lambda}_a$ and fixed $\bar{\beta}=3.16$, chosen so that a wide heterogeneity of behaviors can be observed and $B_0$ lies about the middle of the interval in which $B(k)$ can be varied. {\bf b}. $B(k)$ for fixed $\bar{\lambda}_a=0.4$ and different values of the production scale parameter $\bar{\beta}$.}
\end{figure}

In Fig.~\ref{fig:theo_reduction2}a, we show simulation results of $B(k)$ for different values of $\bar{\lambda}_a$. For all activation rates, a strong inverse dependence of effective burstiness on degree can indeed be observed. For high degree nodes, $B(k)$ is noticeably below $B_0$. This is due to the fact that hubs do not remain in the inactive state for much time since they are usually connected to many active nodes that constantly reactivate them. 

On the other hand, $\bar{\lambda}_a$ still allows to control the duration of inactive periods for low degree nodes. For low values of $\bar{\lambda}_a$ (but above the minimum required to sustain the activity), low degree nodes remain a long time in the inactive state, which raises their effective burstiness well above $B_0$, while for high values of $\bar{\lambda}_a$ they behave as high degree nodes approaching a minimum value of $B(k)$ independent of $k$, a basal level below $B_0$.
Therefore, we conclude that the bursty production of nodes can be regulated by network-driven stochastic activation dynamics that splits the originally uniform endogenous burstiness in a range of values anti-correlated with degree. 

Fixing $\bar{\lambda}_a$, the production shape parameter $\bar{\beta}$ can be varied to regulate the effective level of burstiness, Fig.~\ref{fig:theo_reduction2}b, with similar qualitative behavior. In both cases, the disparity of effective $B(k)$ values reaches a maximum at some intermediate parameter levels. However, $\bar{\beta}$ does not have any limitation to be increased (in our computational experiments it is varied over four orders of magnitude) so that high-degree nodes can increase their effective burstiness above the endogenous level.

\begin{figure}[t]
\includegraphics[width=\linewidth]{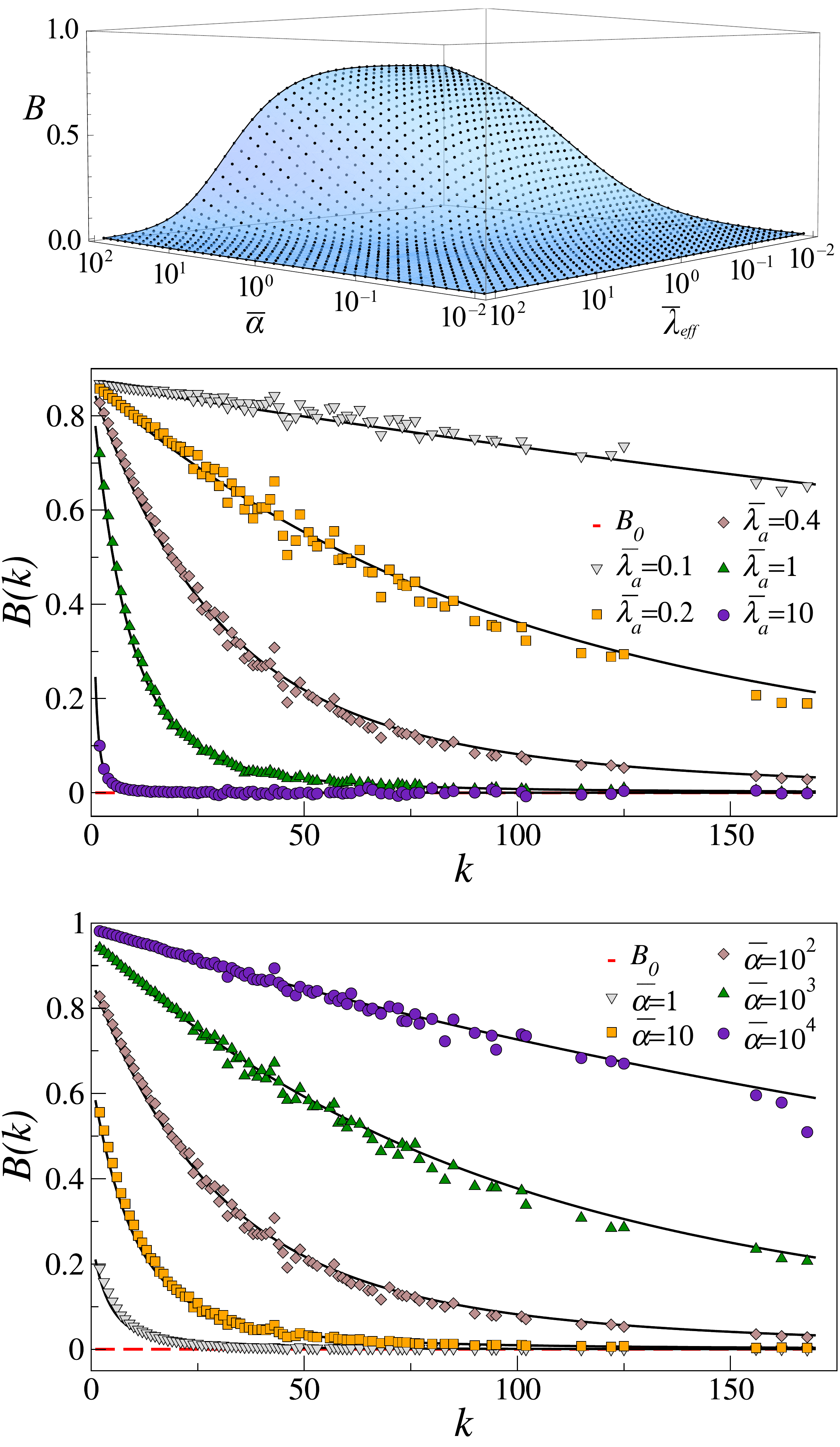}
\caption{\label{fig:theo_induction} {\bf Network-driven induction of burstiness.} In all graphs, solid surface and lines correspond to analytical results and dots to simulations. {\bf a} Comparison between simulation data and the analytical calculation of the effective burstiness $B(\bar{\alpha},\bar{\lambda}_{eff})$, $B>B_{0}=0$. {\bf b}. $B(k)$ for fixed $\bar{\alpha}=100$ and different values of the activation rate $\bar{\lambda}_a$. {\bf c}. $B(k)$ for fixed $\bar{\lambda}_a=0.4$ and different values of the production rate $\bar{\alpha}$.}
\end{figure}

Finally, we also prove that network-driven activation can induce burstiness even when the endogenous production of nodes is not bursty. We let nodes in the active state to produce events with an exponential inter-event time probability distribution function of rate $\alpha$, $\psi(t) = \alpha e^{-\alpha t}$, meaning that the endogenous production of nodes is not bursty, $B_0=0$. Again, calculating the first two moments of Eq.~\eqref{eq:phi} (see Appendix~\ref{sec:phi_poisson}), we obtain
\begin{equation}
CV_\tau(k)=\sqrt{1+\frac{2 \bar{\alpha}}{(1+\bar{\lambda}_{eff}(k))^2}},
\label{bkbi}
\end{equation}
where $\bar{\alpha}$ and $\bar{\lambda}_{eff}(k)$ are the original values rescaled by $\lambda_i$. This analytical expression matches almost perfectly with numerical simulations of a single node whose state changes following Poisson processes with rates $\bar{\lambda}_{eff}$ for activation and $\lambda_i = 1$ for inactivation, and that intrinsically produces events according to an exponential inter-event time probability distribution with rate $\bar{\alpha}$, Fig.~\ref{fig:theo_induction}a. For an exponential distribution, $\psi^{0}(t) = \psi(t)$ so that $\psi^0(t)$ has no role as expected for a Poisson process. Hence, the effective burstiness can only be equal or above the endogenous value $B_0=0$. This induced burstiness is explained as a consequence of the fact that the node may remain in the inactive state for a period of time considerably longer than the average endogenous inter-event time $\langle t_p \rangle= 1/\alpha$~\cite{Dobrzynski:2009}. 

We also simulated exponential production in the same scale-free network with $N=10^4$ nodes and characteristic exponent $\gamma=2.5$ used in the regulation study. We used different values of $\bar{\lambda}_a$ with $\bar{\alpha}=100$, Fig.~\ref{fig:theo_induction}b, and different values of $\bar{\alpha}$ over four orders of magnitude with $\bar{\lambda}_a=0.4$, Fig.~\ref{fig:theo_induction}c. As for bursty production, we find again that $B(k)$ is always a decreasing function of $k$ (but always above $B_0=0$). Due to their sustained activity, hubs are less susceptible to display induced burstiness production. In contrast, nodes with low degree stay in the inactive state for longer periods and are more prone to present a raised effective burstiness. Qualitatively, both parameters $\bar{\alpha}$ and $\bar{\lambda}_a$ can control similarly the induction of bustiness, affecting all nodes more efficiently for higher values of $\bar{\alpha}$ and lower values of $\bar{\lambda}_a$. Again, the disparity of induced burstiness values reaches a maximum at some intermediate parameter levels. 

Different dynamical processes occurring on a network can interplay in unexpected ways. We have seen here that the time series of production events of interacting nodes are affected by network-driven activation/inactivation dynamics. In a broad range of parameters, hubs in a network tend to decrease burstiness in their endogenous pattern of production, hence helping to maintain a stable bulk level, at the expenses of low degree nodes being more pliable to the regulatory effect of the network that tends to amplify their fluctuations. At the same time, lower degree nodes are more prone to burstiness induction. These results indicate that a heterogenous network structure protects the functioning of some nodes, the hubs, making low degree nodes better targets for engineering actions to produce local modifications of production without critically affecting the behavior of the whole system. Taken together, our findings suggest that heterogeneous network interconnectivity may be a strategy in itself developed to protect complex systems against unpredictable functional fluctuations. However, further research should be conducted to determine the effects of different activation/inactivation dynamics on node's burstiness.

\begin{acknowledgements}
This work was supported by a James S. McDonnell Foundation Scholar Award in Complex Systems; the ICREA Academia prize, funded by the {\it Generalitat de Catalunya}; MINECO projects No.\ FIS2013-47282-C2-1, FIS2010-21781-C02-02, and BFU2010-21847-C02-02; {\it Generalitat de Catalunya} grant No.\ 2014SGR608; MICINN Ram\'on y Cajal program, and EC FET-Proactive Project MULTIPLEX (grant 317532).
\end{acknowledgements}

\onecolumngrid
\appendix

\section{Dynamics and notation\label{sec:dynamics}}
The system is composed of a set of elements interconnected forming a scale-free undirected random network. Every node can present two states, active and inactive, and the dynamics that governs their state is similar to the standard SIS model. Nodes flip from one state to the other following a Poisson process with probability densities $\xi_{a}(t)$ for activation and $\xi_{i}(t)$ for deactivation. In particular, a node $l$ flips from the active to the inactive state at a constant rate $\lambda_{i}$, while its activation rate is proportional to the number of active nodes to which it is connected, $\lambda_{a,l}(t)=\lambda_{a} \sum_{m} a_{lm}n_{m}(t)$. In the latter sum, $n_{m}(t)=1$ if node $m$ is active at time $t$ and $n_{m}(t)=0$ otherwise; similarly, $a_{lm}=1$ if and only if nodes $l$ and $m$ are connected.

We now show that $\lambda_{a,l}(t)$ is, after the system has reached the steady-state, approximately time-independent and proportional to the degree of the node $k_{l}$~\cite{Pastor-Satorras:2001}; we will refer to it as $\lambda_{eff}(k)$. To derive its expression, we first apply a coarse-graining approximation by assuming that all nodes with the same degree $k$ are equivalent and statistically independent. Under this approximation, the state of the system is characterized by the fraction of nodes in the active state in every degree class,
\begin{equation}
\rho_{k}(t) = \frac{\sum \limits_{i | k_{i} = k} n_{i}(t)}{N_{k}},
\end{equation}
where $N_{k}$ stands for the number of nodes with degree $k$ in the network. Since we are assuming all nodes with the same degree to be statistically independent, $\rho_{k}(t)$ yields the probability that a randomly chosen node with degree $k$ is active. Hence, the probability that a randomly chosen link emerging from a node with degree $k$ reaches an active node can be computed as $\sum_{k'} P(k'|k) \rho_{k'}(t)$, i.e., the probability that the link reaches a node with degree $k'$, $P(k'|k)$, and that such node is active, $\rho_{k'}(t)$, summed for all $k'$. In the case of uncorrelated random networks, $P(k'|k) = k' \frac{P(k')}{\langle k \rangle}$, so the probability for a link to connect to an active node becomes degree independent, $\rho(t) \equiv \sum_{k'} \rho_{k'}(t) k' \frac{P(k')}{\langle k \rangle}$. Moreover, in the steady-state it takes an almost constant value $\bar{\rho} \equiv \left\langle \sum_{k'} \rho_{k'}(t) k' \frac{P(k')}{\langle k \rangle} \right\rangle$ (see Fig.~\ref{fig:wprevalence}), which allows us to compute the average number of active neighbors of a node with degree $k$ as $k \bar{\rho}$. Hence,
\begin{equation}\label{eq:act_rate}
\lambda_{eff}(k) = \lambda_{a} k \bar{\rho} = \lambda_{a} k \left\langle \sum_{k'} \rho_{k'}(t) k' \frac{P(k')}{\langle k \rangle} \right\rangle.
\end{equation}

In addition to the SIS-like dynamics, we let nodes produce while they are in the active state with a probability density $\psi(t)$, which need not be exponential. It is therefore important to note that, if $\psi(t) \text{d}t$ gives the probability for the time interval between two consecutive production events to fall into $\left( t,t + \text{d}t \right)$ (for a node that is active along all the process), the probability density for the interval between the activation time and the first production event $\psi^{0}(t)$ is, in general, given by
\begin{equation}\label{eq:psi_0}
\psi^{0}(t) = \frac{\Psi(t)}{\langle t_{p} \rangle},
\end{equation}
where $\langle t_{p} \rangle = \int \limits_{0}^{\infty} t \psi(t)\text{d}t$ is the average production interval and $\Psi(t) = \int \limits_{t}^{\infty} \psi(t')\text{d}t'$ is the survival probability distribution (the probability that the interval is greater than $t$). We will use the same notation for the activation and deactivation probabilities as well: $\Xi_{a/i}(t)$ will refer to the survival probability of $\xi_{a/i}(t)$. To justify Eq.~\eqref{eq:psi_0}, let us consider a sequence of $N \to \infty$ events $t_{i}, i=1,\ldots,N$ distributed with probability density $\psi(t)$ (so the whole process takes a time $T = \sum_{i=1}^{N} t_{i}$) and calculate $\psi^{0}(t) \text{d}t$, i.e., the probability that an activation event falls at a distance between $t$ and $t+\text{d}t$ from the next production event. Assuming that the activation event is uniformly distributed along the interval $(0,T)$, the probability for it to fall in any interval of length $L$ is simply $\frac{L}{T}$. In our case, $L = N_{t_{i} > t} \text{d}t$, where $N_{t_{i}>t}$ is the number of intervals greater than $t$, since there is an interval of length $\text{d}t$ placed at distance $t$ from the next production event in every production interval greater than $t$. Therefore,
\begin{equation}
\psi^{0}(t) \text{d}t = \lim \limits_{N \to \infty} \frac{N_{t_{i} > t} \text{d}t}{T} = \lim \limits_{N \to \infty} \frac{\frac{N_{t_{i} > t}}{N}}{\frac{\sum\limits_{i=1}^{N} t_{i}}{N}} \text{d}t = \frac{\Psi(t)}{\langle t_{p} \rangle} \text{d}t.
\end{equation}

\section{Probability density function $\phi(t)$\label{sec:phi}}
In the latter section we have reduced the study of the system to that of a single node that activates with rate $\lambda_{eff}(k)$ given by Eq.~\eqref{eq:act_rate}, deactivates with rate $\lambda_{i}$ and produces in the active state with probability densities $\psi(t)$ and $\psi^{0}(t)$. We now compute the probability $\phi(t)\text{d}t$ that two consecutive production events take place in the interval $\left( t, t + \text{d}t \right)$ taking into account the fact that the state of the node can flip any number of times between the two events (as long as it is in the active state during both productions). It is easy to see that if it does not flip to the inactive state during the process, $\phi(t) \text{d}t$ is given by $\Xi_{i}(t) \psi(t) \text{d}t$ (the probability that it does not deactivate in a time less than $t$ and that it produces in the given interval). Similarly, we can compute the probability that it deactivates only once between both productions; however, in this case we have to integrate for all possible deactivation and activation times (let us call them $t_{1}$ and $t_{2}$, respectively), yielding $\int\limits_{0}^{t} \int\limits_{0}^{t-t_{1}} \Psi(t_{1}) \xi_{i}(t_{1}) \xi_{a}(t_{2}) \psi^{0}(t-t_{1}-t_{2}) \Xi_{i}(t-t_{1}-t_{2}) \text{d}t_{1} \text{d}t_{2} \text{d}t$. Following the same procedure, we can write the total probability density as the series
\begin{equation}\label{eq:phi_sup}
\begin{aligned}
\phi(t) &= \Xi_{i}(t) \psi(t) + \int\limits_{0}^{t} \int\limits_{0}^{t-t_{1}} \Psi(t_{1}) \xi_{i}(t_{1}) \xi_{a}(t_{2}) \psi^{0}(t-t_{1}-t_{2}) \Xi_{i}(t-t_{1}-t_{2}) \text{d}t_{1} \text{d}t_{2} \\
&+ \int\limits_{0}^{t} \int\limits_{0}^{t-t_{1}} \int\limits_{0}^{t-t_{1}-t_{2}} \int\limits_{0}^{t-t_{1}-t_{2}-t_{3}} \Psi(t_{1}) \xi_{i}(t_{1}) \xi_{a}(t_{2}) \xi_{i}(t_{3}) \Psi^{0}(t_{3}) \xi_{a}(t_{4}) \\
& \times \psi^{0}(t-t_{1}-t_{2}-t_{3}-t_{4}) \Xi_{i}(t-t_{1}-t_{2}-t_{3}-t_{4}) \text{d}t_{1} \text{d}t_{2} \text{d}t_{3} \text{d}t_{4} + \ldots
\end{aligned}
\end{equation}
We do not need to solve the series, since we are only interested in the first two moments of the distribution to obtain $B(k)$. To this end, we make use of the Laplace transform formalism, which is particularly convenient for two reasons. On the one hand, the $n$-th moment of a distribution $\phi(t)$ can be obtained as the $n$-th derivative of its transform $\hat{\phi}(s)$, $\langle t^{n} \rangle = (-1)^{n} \left. \hat{\phi}^{(n)}(s) \right|_{s=0}$. On the other hand, the transform of a convolution is $ \mathcal{L} \left\{ ( f \star g ) (t) \right\} = \hat{f}(s) \cdot \hat{g}(s)$. Since any term in Eq.~\eqref{eq:phi_sup} is a chain of convolutions, it is not hard to write down $\hat{\phi}(s)$,
\begin{equation}
\begin{aligned}
\hat{\phi}(s) &= \mathcal{L}\left\{ \Xi_{i}(t) \psi(t) \right\} + \mathcal{L}\left\{ \Psi(t) \xi_{i}(t) \right\} \hat{\xi}_{a}(s)\mathcal{L}\left\{ \psi^{0}(t) \Xi_{i}(t) \right\} \\
&+ \mathcal{L}\left\{ \Psi(t) \xi_{i}(t) \right\} \hat{\xi}_{a}(s) \mathcal{L}\left\{ \Psi^{0}(t) \xi_{i}(t) \right\} \hat{\xi}_{a}(s) \mathcal{L}\left\{ \psi^{0}(t) \Xi_{i}(t) \right\} + \ldots \\
&= \mathcal{L}\left\{ \psi(t) \Xi_{i}(t) \right\} + \frac{\mathcal{L}\left\{ \Psi(t) \xi_{i}(t) \right\} \mathcal{L}\left\{ \psi^{0}(t) \Xi_{i}(t) \right\} }{\mathcal{L}\left\{ \Psi^{0}(t) \xi_{i}(t) \right\}} \sum \limits_{n=1}^{\infty} \left( \hat{\xi}_{a}(s) \mathcal{L}\left\{ \Psi^{0}(t) \xi_{i}(t) \right\} \right)^{n} \\
&= \mathcal{L}\left\{ \psi(t) \Xi_{i}(t) \right\} + \mathcal{L}\left\{ \Psi(t) \xi_{i}(t) \right\} \mathcal{L}\left\{ \psi^{0}(t) \Xi_{i}(t) \right\} \frac{\hat{\xi}_{a}(s)}{1 - \hat{\xi}_{a}(s)\mathcal{L}\left\{ \Psi^{0}(t) \xi_{i}(t) \right\}},
\end{aligned}
\end{equation}
where in the last step we have made use of the fact that $(1-\alpha) \sum_{n=1}^{\infty} \alpha^{n} = \alpha$. We can write the latter result in terms of $\psi(t)$ only by plugging the expressions for $\xi_{a}(t)$ and $\xi_{i}(t)$ into it and using
\begin{equation}
\mathcal{L} \left\{ f(t) e^{- \alpha t} \right\} = \hat{f}(s + \alpha),
\end{equation}
\begin{equation}
\hat{\Psi}(s) = \mathcal{L} \left\{ \int \limits_{t}^{\infty} \psi(t') \text{d}t' \right\} = \mathcal{L} \left\{ 1 - \int \limits_{0}^{t} \psi(t') \text{d}t' \right\} = \frac{1}{s} \left( 1 - \hat{\psi} (s) \right)
\end{equation}
and
\begin{equation}
\hat{\psi}^{0}(s) = \mathcal{L} \left\{ \frac{\Psi(t)}{\langle t _{p}\rangle} \right\} = \frac{1}{\langle t _{p}\rangle} \frac{1}{s} \left( 1 - \hat{\psi} (s) \right).
\end{equation}
The final expression for $\hat{\phi}(s)$ reads
\begin{equation}\label{eq:phi_p}
\hat{\phi}(s) = \hat{\psi}(s+\lambda_{i}) + \frac{\frac{\lambda_{i}}{\langle t _{p}\rangle \left( s + \lambda_{i} \right)^{2}} \left[ 1 - \hat{\psi} (s + \lambda_{i}) \right]^{2}}{\frac{s + \lambda_{eff}(k)}{\lambda_{eff}(k)} - \frac{\lambda_{i}}{s + \lambda_{i}} \left[ 1 - \frac{1 - \hat{\psi} (s + \lambda_{i})}{\langle t _{p}\rangle \left( s + \lambda_{i} \right)} \right]}.
\end{equation}
Up to this point, we have solved the general problem for an unspecified production probability density $\psi(t)$. In this paper we have considered two different cases, which will be studied separately in two subsections.

\subsection{Burstiness regulation for a Weibull density function\label{sec:phi_weibull}}
In the first stage of this work, we have considered a Weibull probability density function with scale parameter $\frac{1}{2 \beta}$ and shape parameter $\frac{1}{2}$,
\begin{equation}\label{eq:weib_prod}
\psi(t) = \sqrt{\frac{\beta }{2 t}} e^{-\sqrt{2 \beta  t}},
\end{equation}
whose intrinsic burstiness coefficient is $B_{0} = \frac{\sqrt{5}-1}{\sqrt{5}+1} \approx 0.382$. Plugging its Laplace transform
\begin{equation}
\hat{\psi} (s) = \sqrt{\frac{\pi }{2}} e^{\frac{\beta }{2 s}} \sqrt{\frac{\beta }{s}} \text{erfc}\left(\sqrt{\frac{\beta }{2s}}\right)
\end{equation}
into Eq.~\eqref{eq:phi_p} yields a rather cumbersome expression,
\begin{equation}
\begin{aligned}
\hat{\phi}(s) &= e^{\frac{\beta}{2 \left( s + \lambda_{i} \right)}} \sqrt{\frac{\pi \beta}{2 \left( s + \lambda_{i} \right)}} \text{erfc} \left( \sqrt{\frac{\beta}{2 \left( s + \lambda_{i} \right)}} \right) \\
&+ \frac{\lambda_{i} \lambda_{eff}(k) \beta \left[ e^{\frac{\beta}{2 \left( s + \lambda_{i} \right)}} \sqrt{\frac{2 \pi \beta}{s + \lambda_{i} }} \text{erfc} \left( \sqrt{\frac{\beta}{2 \left( s + \lambda_{i} \right)}} \right) - 2 \right]^{2}}{4 s \left( s + \lambda_{i} \right) \left( s + \lambda_{i} + \lambda_{eff}(k) \right) + 4 \lambda_{i} \lambda_{eff}(k) \beta - 2 \lambda_{i} \lambda_{eff}(k) \beta e^{\frac{\beta}{2 \left( s + \lambda_{i} \right)}} \sqrt{\frac{2 \pi \beta}{s + \lambda_{i} }} \text{erfc} \left( \sqrt{\frac{\beta}{2 \left( s + \lambda_{i} \right)}} \right)}.
\end{aligned}
\end{equation}
We can now obtain the first and second moments of $\phi(t)$ deriving the latter equation and therefore write the expressions for the mean $\mu_{\tau}(k) = - \left. \hat{\phi}^{(1)}(s) \right|_{s=0}$, the standard deviation $\sigma_{\tau}(k) = \sqrt{\left. \hat{\phi}^{(2)}(s) \right|_{s=0} - \mu_{\tau}^2(k)}$ and $B(k) = \left( \sigma_{\tau}(k) - \mu_{\tau}(k) \right)/\left( \sigma_{\tau}(k) + \mu_{\tau}(k) \right)$ in terms of the normalized parameters $\bar{\beta} \equiv \frac{\beta}{\lambda_{i}}$ and $\bar{\lambda}_{a} \equiv \frac{\lambda_{a}}{\lambda_{i}}$:
\begin{equation}\label{eq:mu_reg}
\mu_{\tau}(k) = \frac{1 + k \bar{\rho } \bar{\lambda }_a}{k \bar{\rho } \bar{\lambda }_a \bar{\beta } \lambda_i },
\end{equation}
\begin{equation}
\sigma_{\tau}(k) = \frac{1 + k \bar{\rho } \bar{\lambda }_a}{k \bar{\rho } \bar{\lambda }_a \bar{\beta } \lambda_i } \sqrt{\frac{4}{2 - \sqrt{2 \pi \bar{\beta}}e^{\frac{\bar{\beta}}{2}}\text{erfc}\left( \sqrt{\frac{\bar{\beta}}{2}} \right)} - 1 + 2 \bar{\beta} \left( \frac{1}{\left( 1 + k \bar{\rho} \bar{\lambda}_{a} \right)^{2}} - 1 \right)}.
\end{equation}
Hence,
\begin{equation}
B(k) = \frac{\sqrt{\frac{4}{2 - \sqrt{2 \pi \bar{\beta}}e^{\frac{\bar{\beta}}{2}}\text{erfc}\left( \sqrt{\frac{\bar{\beta}}{2}} \right)} - 1 + 2 \bar{\beta} \left( \frac{1}{\left( 1 + k \bar{\rho} \bar{\lambda}_{a} \right)^{2}} - 1 \right)}-1}{\sqrt{\frac{4}{2 - \sqrt{2 \pi \bar{\beta}}e^{\frac{\bar{\beta}}{2}}\text{erfc}\left( \sqrt{\frac{\bar{\beta}}{2}} \right)} - 1 + 2 \bar{\beta} \left( \frac{1}{\left( 1 + k \bar{\rho} \bar{\lambda}_{a} \right)^{2}} - 1 \right)}+1}.
\end{equation}
\subsection{Burstiness regulation for an exponential density function\label{sec:phi_poisson}}
We can repeat the same procedure as in the latter subsection for an exponential production probability density,
\begin{equation}
\psi(t) = \alpha e^{- \alpha t},
\end{equation}
whose intrinsic burstiness coefficient is $B_{0}=0$. Its Laplace transform is given by
\begin{equation}
\hat{\psi}(s) = \frac{\alpha}{\alpha + s},
\end{equation}
so Eq.~\eqref{eq:phi_p} becomes
\begin{equation}
\hat{\phi}(s) = \frac{\alpha(\lambda_{eff}(k)+s)}{s (\lambda_{i}+\lambda_{eff}(k)+s)+\alpha(\lambda_{eff}(k)+s)}.
\end{equation}
Evaluating its first and second derivatives at $s=0$ and using normalized parameters $\bar{\alpha} \equiv \frac{\alpha}{\lambda_{i}}$ and $\bar{\lambda}_{a} \equiv \frac{\lambda_{a}}{\lambda_{i}}$ gives
\begin{equation}\label{eq:mu_ind}
\mu_{\tau}(k) = \frac{1 + k \bar{\rho } \bar{\lambda }_a}{k \bar{\rho } \bar{\lambda}_a \bar{\alpha } \lambda_i },
\end{equation}
\begin{equation}
\sigma_{\tau}(k) = \frac{1 + k \bar{\rho } \bar{\lambda }_a}{k \bar{\rho } \bar{\lambda}_a \bar{\alpha } \lambda_i } \sqrt{1 + \frac{2 \bar{\alpha}}{\left( 1 + k \bar{\rho} \bar{\lambda}_{a} \right)^{2}}}
\end{equation}
and
\begin{equation}
B(k) = \frac{\sqrt{1 + \frac{2 \bar{\alpha}}{\left( 1 + k \bar{\rho} \bar{\lambda}_{a} \right)^{2}}}-1}{\sqrt{1 + \frac{2 \bar{\alpha}}{\left( 1 + k \bar{\rho} \bar{\lambda}_{a} \right)^{2}}}+1}.
\end{equation}

\section{Limiting cases and scope of the results\label{sec:limits}}
Let us focus on the case of a single node that produces with an arbitrary $\psi(t)$ and consider several limiting cases. In order to define a timescale, we set $\beta \equiv \left\langle t_{p} \right\rangle^{-1} = 1$ throughout all this section without loss of generality (so the ratios $\lambda_{i}/\beta$ and $\lambda_{eff}/\beta$ to which we will refer become simply $\lambda_{i}$ and $\lambda_{eff}$).

In the limit of $\lambda_{i} \to 0$, the fraction of production intervals interrupted by a deactivation goes to zero, so the resulting probability density function $\phi(t) \to \psi(t)$. This result can also be derived by taking the limit of Eq.~\eqref{eq:phi_p} as $\lambda_{i} \to 0$, since $\hat{\phi}(s) \to \hat{\psi}(s)$. Notice, however, that this limit is slightly different to the limit $\beta/\lambda_{i} \to \infty$ and fixed $\lambda_{eff}/\lambda_{i}$ (which gives $B(k) \to 1$); the reason is that, in the present case, when taking the limit $\lambda_{i}/\beta \to 0$ for a fixed $\lambda_{eff}/\beta$, the ratio $\beta/\lambda_{eff}$ obviously remains constant, i.e. the interrupted intervals become less frequent but not longer than uninterrupted ones. On the other hand, taking the limit $\beta/\lambda_{i} \to \infty$ for a fixed $\lambda_{eff}/\lambda_{i}$ gives $\beta/\lambda_{eff} \to \infty$. Therefore, even though interrupted intervals become less frequent, they become much longer than uninterrupted intervals as well, hence yielding a bursty behavior. That is why, when working with the normalized parameters $\bar{\beta} = \beta/\lambda_{i}$ and $\bar{\lambda}_{eff} = \lambda_{eff}/\lambda_{i}$, $\bar{\lambda}_{eff}$ must be increased too in order to recover the original burstiness coefficient $B_{0}$ (as shown in Fig.~\ref{fig:theo_reduction1}).

For intermediate values of the deactivation rate $0 < \lambda_{i} < 1$, the fraction of production events interrupted by a deactivation may be significant; there is, on average, one production event per unit time during the active state, while there are $\lambda_{i}$ interruptions per unit time. Therefore, approximately one out of every $\lambda_{i}^{-1}$ production intervals is interrupted. In addition, the average time the node remains in the inactive state is $\lambda_{eff}(k)^{-1}$, so if $\lambda_{eff}(k) \ll 1$, $\lambda_{i}^{-1}$ production events buffer during the active state according to $\psi(t)$ and are followed by a long period of inactivity. Hence, the resulting distribution becomes burstier than the original.

In the opposite case ($\lambda_{i} \to \infty$), the production process becomes Poisson regardless of $\psi(t)$; since $\langle t_{p}^{0} \rangle \gg \langle t_{i} \rangle$, the probability that the node produces during an active period (which we will denote by $\nu$) is very small, and we can regard the process as a succession of active and inactive states of average length $\langle t_{a+i} \rangle = \frac{1}{\lambda_{eff}(k)} + \frac{1}{\lambda_{i}}$, each of which with an independent production probability $\nu$. Under this approximation, we can easily compute the probability that two consecutive production events take place in a time interval $T$ greater than $t$ as the probability for the node not to produce in any of $\frac{t}{\langle t_{a+i} \rangle}$ intervals,
\begin{equation}
\text{Prob} \lbrace T > t \rbrace = \left( 1 - \nu \right)^{\frac{t}{\langle t_{a+i} \rangle}} = e^{\frac{t}{\langle t_{a+i} \rangle} \ln \left( 1 - \nu \right)} \approx e^{- \frac{\nu}{\langle t_{a+i} \rangle} t} = e^{- \frac{\lambda_{i} \lambda_{eff}(k) \nu}{\lambda_{i} + \lambda_{eff}(k)} t}.
\end{equation}
Furthermore, $\nu$ can be estimated as well for any $\psi(t)$. Its exact expression is given by
\begin{equation}\label{eq:rho1}
\nu = \int \limits_{0}^{\infty} \psi^{0} (t) \Xi_{i} (t) \text{d}t = \int \limits_{0}^{\infty} \Psi (t) e^{- \lambda_{i} t} \text{d}t.
\end{equation}
Since we are considering the limit of $\langle t_{p}^{0} \rangle \gg \langle t_{i} \rangle$, the probability that a production interval is greater than $\langle t_{i} \rangle$ is approximately one, so $\Psi (t) \sim 1$ in the range of small values of $t$ in which the exponential term in the integrand is not close to zero. Hence, we can approximate Eq.~\eqref{eq:rho1} as
\begin{equation}
\nu \sim \int \limits_{0}^{\infty} e^{- \lambda_{i} t} \text{d}t = \frac{1}{\lambda_{i}}.
\end{equation}
We therefore see that the production process becomes exponentially distributed with mean
\begin{equation}\label{eq:mean_limit}
\mu_{\tau} (k) = \frac{\lambda_{i} + \lambda_{eff}(k)}{\lambda_{eff}(k)},
\end{equation}
which is in accordance with Eqs.~(\ref{eq:mu_reg},\ref{eq:mu_ind}). This result can also be obtained by taking the limit of $\hat{\phi}(s)$ as $\lambda_{i} \to \infty$ assuming
\begin{equation}\label{eq:laplace_limit}
\lim \limits_{\lambda_{i} \to \infty} \hat{\psi}(s+\lambda_{i}) = \lim \limits_{s \to \infty} \hat{\psi}(s) = \lim \limits_{s \to \infty} \int \limits_{0}^{\infty} e^{-st} \psi(t) \text{d}t = 0
\end{equation}
and
\begin{equation}
\lim \limits_{\lambda_{i} \to \infty} \Psi(s+\lambda_{i}) = \lim \limits_{\lambda_{i} \to \infty} \frac{1}{s+\lambda_{i}} \left( 1 - \psi(s+\lambda_{i}) \right) = \lim \limits_{\lambda_{i} \to \infty} \frac{1}{s + \lambda_{i}}.
\end{equation}
From all the stated above, we can conclude that the network dynamics can be used to control the production of nodes between three regimes of special interest: the endogenous production behavior, a bursty process and a Poisson process (see Table~\ref{tab:behaviours}). Furthermore, these results are completely general and independent of the functional form of $\psi(t)$ (as long as Eq.~\eqref{eq:laplace_limit} holds).

\begin{table}[h]
\begin{tabular}{c|c|c}
Parameters range & $\phi(t)$ & $B$ \\ \hline
$\lambda_{i}/\beta \to 0$ and fixed $\lambda_{eff}(k)/\beta$ & $\psi(t)$ & $B_{0}$ \\
$0 < \lambda_{i}/\beta < 1, \lambda_{eff}(k)/\beta \ll 1$ & Bursty distribution & $>B_{0}$ \\
$\lambda_{i}/\beta \to \infty$ & $\frac{\lambda_{eff}(k)\beta}{\lambda_{i} + \lambda_{eff}(k)} e^{-\frac{\lambda_{eff}(k)\beta}{\lambda_{i} + \lambda_{eff}(k)} t}$ & 0
\end{tabular}
\caption{\label{tab:behaviours}Different behaviors exhibited by a node following the dynamics presented in the paper.}
\end{table}
\section{Simulation details\label{sec:simulation}}
The IAI dynamics is simulated using the continuous-time Gillespie algorithm. In addition, at every deactivation event the deactivated node's production is updated by generating random intervals with distributions $\psi^{0}(t)$ for the first one and $\psi(t)$ for the rest of them. The initial condition for all simulations is $n_{i}(0) = 1, \, \forall i$ (all nodes active), but we only measure nodes' $B$ for simulation times $t > 100$, since the system reaches a stationary state before that value (see Fig.~\ref{fig:wprevalence}). We stop simulations when all nodes have produced at least $10^{4}$ times, so enough statistics are guaranteed.

In Figs.~2 and 3 in the paper, the simulation parameters have been chosen in such a way that a wide heterogeneity of behaviours is exhibited; in Figs.~2a and 3b, every simulation corresponds to a different value of $\bar{\lambda}_{a}$ between $\bar{\lambda}_{a} = 0.1$ and $\bar{\lambda}_{a} = 10$, so $\bar{\rho}$ takes values approximately between $\bar{\rho} \approx 0.1$ and $\bar{\rho} \approx 1$. Since the degree of nodes is between $k = 1$ and $k \approx 170$, $\bar{\lambda}_{eff}(k)$ is approximately in the intervals $(0.01,1.7)$ for $\bar{\lambda}_{a}=0.1$ and $(10,1700)$ for $\bar{\lambda}_{a}=10$. From Fig.~1 it can be seen that for $\bar{\beta} \approx 3$, $B_{0}$ lies about the middle of the interval in which $B$ can be varied; we have set $\bar{\beta}=3.16$. A similar criterion has been applied when setting $\bar{\alpha}=100$. Likewise, in Figs.~2b and 3c $\bar{\lambda}_{a}=0.4$ is fixed, so $\bar{\rho} \approx 0.54$ and hence $\bar{\lambda}_{eff}(k) \in (0.2, 37)$. Given that $\bar{\alpha}$ and $\bar{\beta}$ are varied over 4 orders of magnitude, a wide range of $B$ values is obtained.

\section{Supplementary results\label{sec:results}}
\begin{figure}[t!]
\includegraphics[width=0.5\columnwidth]{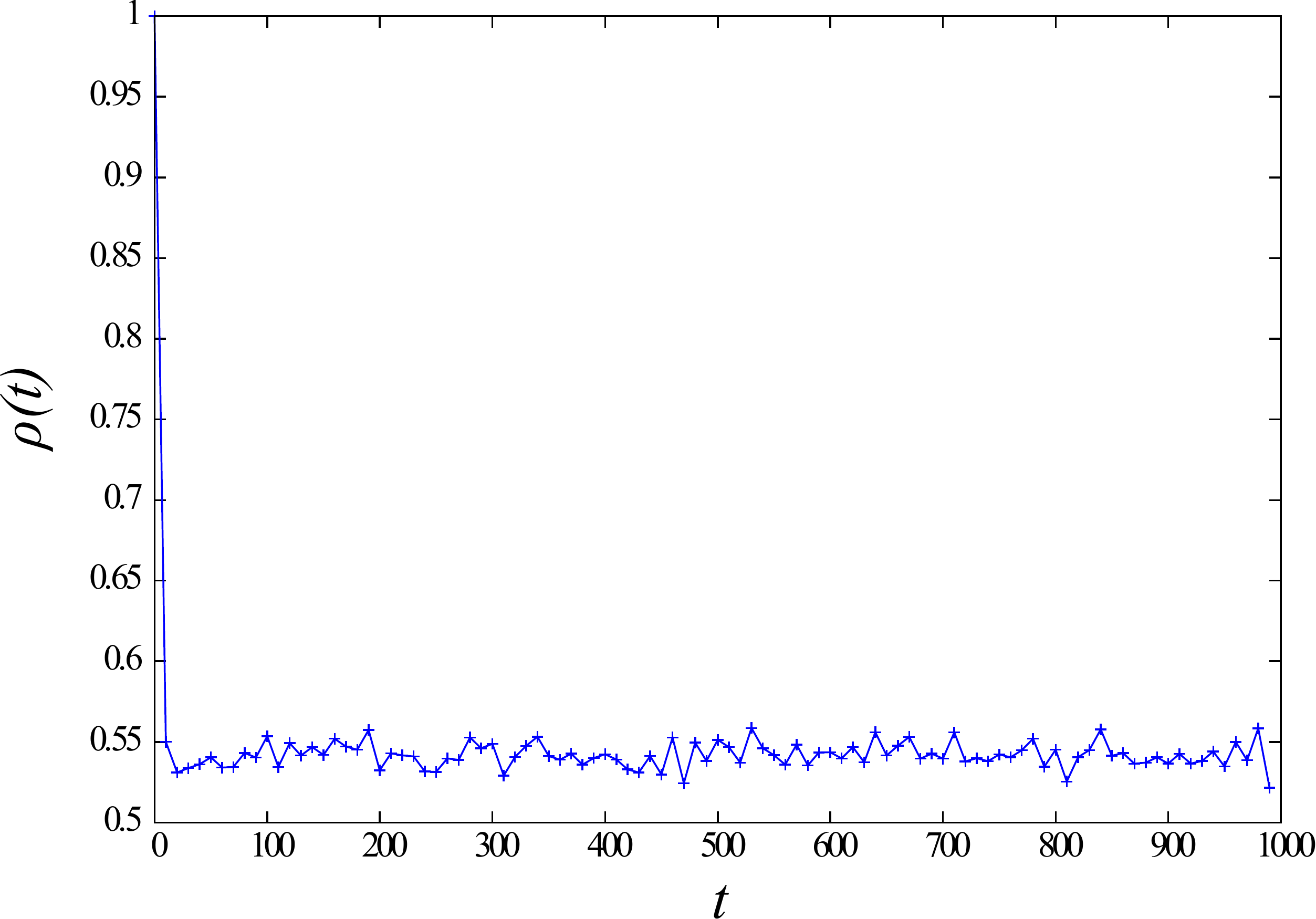}
\caption{\label{fig:wprevalence} Probability for a link to be active as a function of time $\rho(t)$.}
\end{figure}

In this section we present some results not included in the paper. In Fig.~\ref{fig:wprevalence} we show the quantity $\rho(t) = \sum_{k'} \rho_{k'}(t) k' \frac{P(k')}{\langle k \rangle}$ derived in Section~\ref{sec:dynamics}. As expected, this quantity fluctuates around a constant value that can be easily measured from simulations.

\begin{figure}[t]
\includegraphics[width=0.45\textwidth]{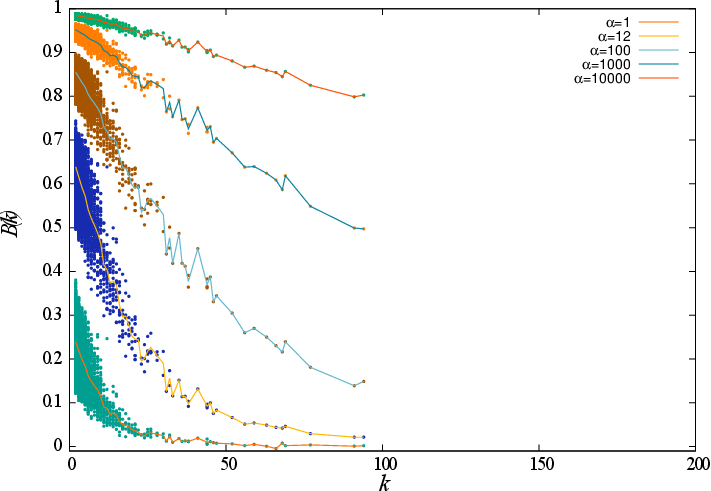}
\includegraphics[width=0.45\textwidth]{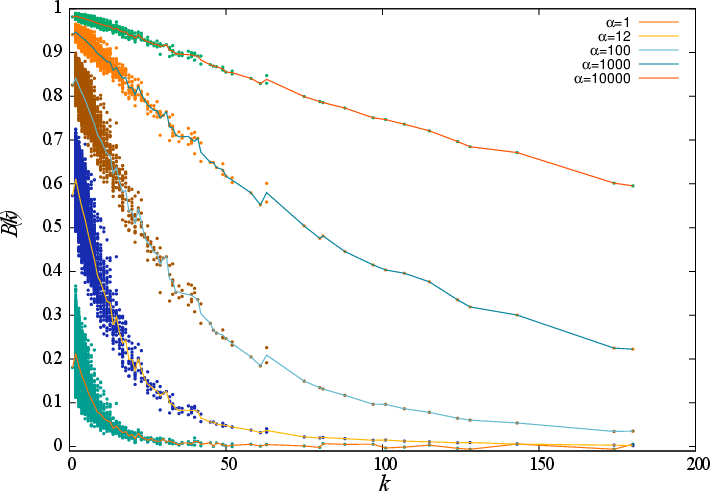}
\includegraphics[width=0.45\textwidth]{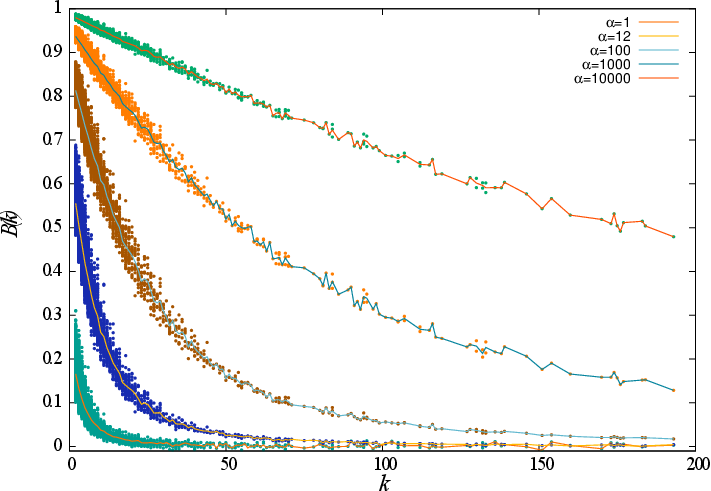}
\includegraphics[width=0.45\textwidth]{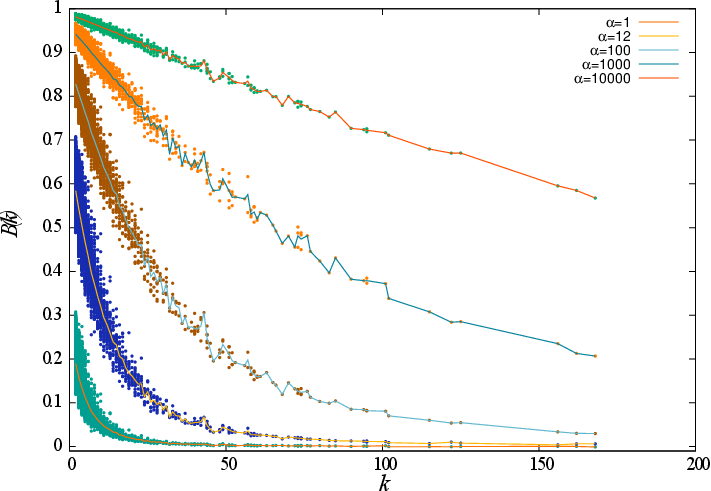}
\caption{\label{fig:heterogeneity} $B(k)$ for SF graphs with different exponents $\gamma$. Top-left: $\gamma=3$. Top-right: $\gamma=2.75$. Bottom-left: $\gamma=2.25$. Bottom-right: $\gamma=2.5$.}
\vspace*{0.5cm}
\includegraphics[width=0.45\textwidth]{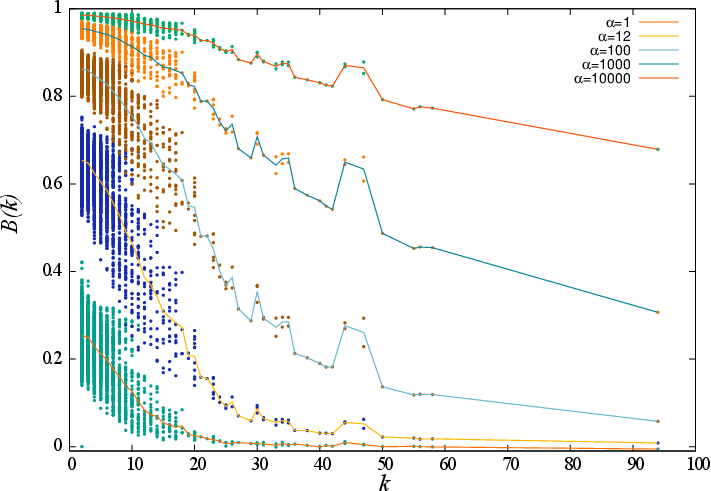}
\includegraphics[width=0.45\textwidth]{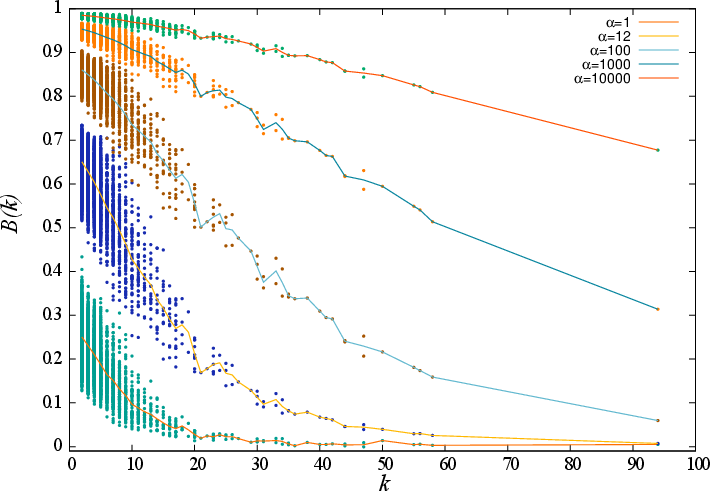}
\includegraphics[width=0.45\textwidth]{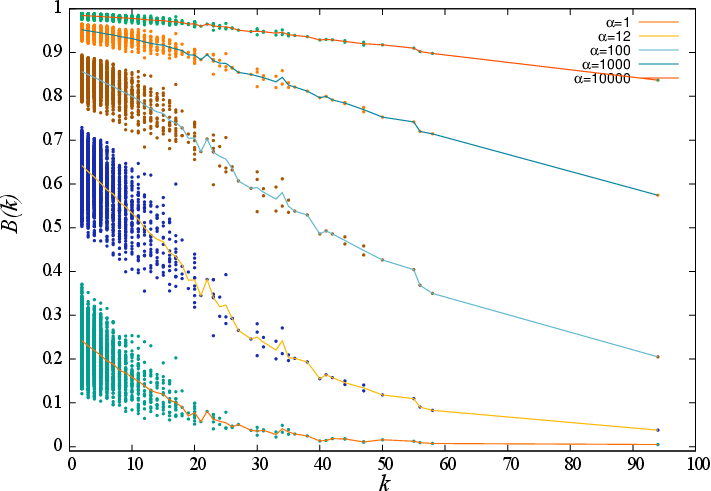}
\includegraphics[width=0.45\textwidth]{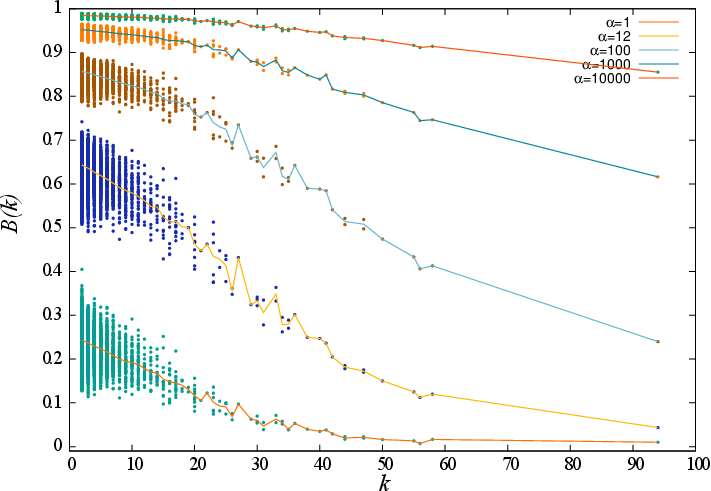}
\caption{\label{fig:assortativity} $B(k)$ for SF graphs with different assortativity $r$. Top-left: $r=1.0$. Top-right: $r=0.5$. Bottom-left: $r=-0.5$. Bottom-right: $r=-1.0$.}
\end{figure}

To see how the network topology affects the induction of burstiness, we have measured $B(k)$ for different networks. In Fig.~\ref{fig:heterogeneity} we show the results for four SF networks with  $N=10^{3}$, $\left\langle k \right\rangle \approx 4$ and $\bar{\lambda}_{a} = 0.4$, but different exponents. In Fig.~\ref{fig:assortativity}, the results correspond to four SF random graphs with the same degree distribution but different assortativities. Our results show a weak dependence on degree heterogeneity; indeed, the higher the heterogeneity, the faster the decay of $B(k)$ with $k$, although the effect is not really significant. Similarly, degree correlations slightly affect the burstiness of high degree nodes, since it is greater for disassortative networks. Yet, assortativity does not play an important role at burstiness induction either.

\begin{figure}[h!]
\includegraphics[width=0.45\columnwidth]{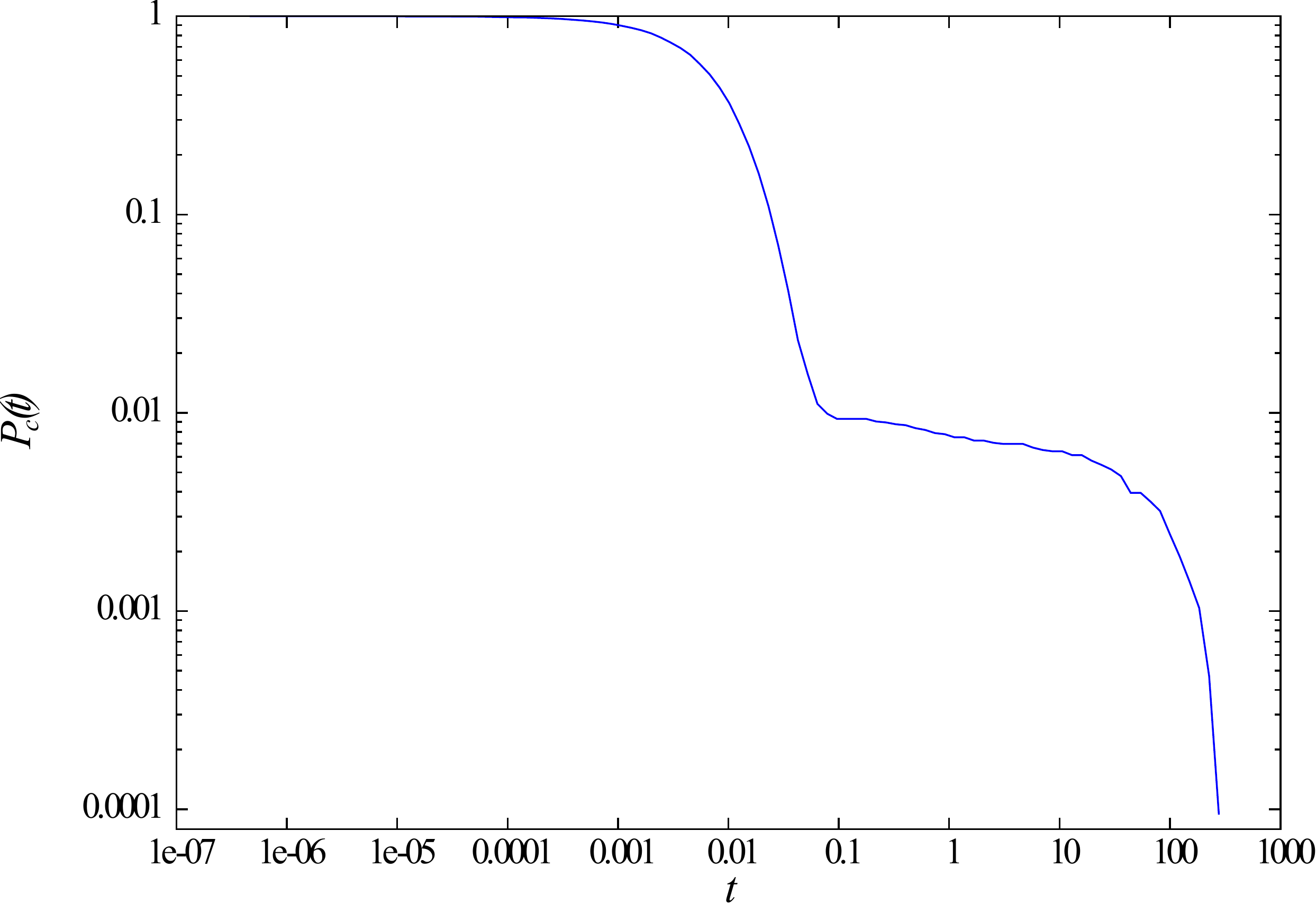}
\includegraphics[width=0.45\columnwidth]{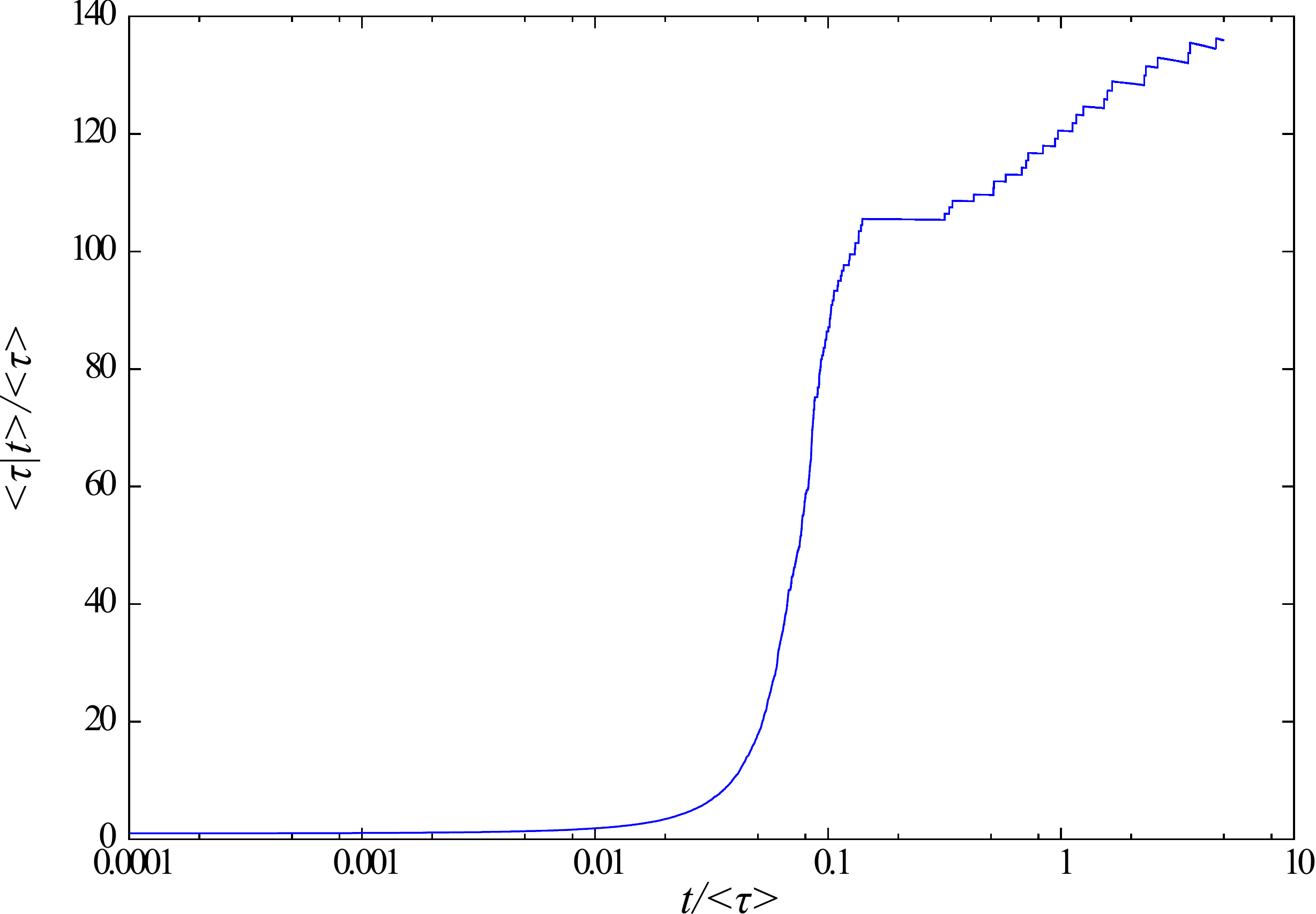}
\caption{\label{fig:burstiest} Left: Cumulative inter-event time distribution function of the node with highest $B$ in the network (for $\bar{\alpha}=100$). Right:  Normalized conditional average interval $\langle \tau | t \rangle/\langle \tau \rangle$ as a function of $t/\langle \tau \rangle$. We see that it clearly increases above the constant value of 1 expected for a non-bursty Poisson process.}
\end{figure}

Since the burstiness coefficient $B(k)$ alone does not provide any details about the distribution other than the ratio between $\sigma_{\tau}$ and $\mu_{\tau}$, we have measured the cumulative inter-event time distribution function for the burstiest node in the network, as shown in Fig.~\ref{fig:burstiest} (left). We have also measured the burstiness of the same distribution by computing the normalized conditional average interval $\langle \tau | t \rangle/\langle \tau \rangle$ as a function of $t/\langle \tau \rangle$ (see Fig.~\ref{fig:burstiest}, right). For a Poisson process, the conditional average interval $\langle \tau | t \rangle$ is independent of $t$ and equal to $\langle \tau \rangle$, while for a regular-like distribution, this should be a decreasing function (the more time elapsed since the last event, the less time is expected until the next one). In our case, we observe exactly the opposite situation, which evidences the counter-intuitive behaviour of bursty distributions; the more time elapsed since the last event, the more time expected until the next one. We thus see that the network clearly induces burstiness on the activity of nodes.

\FloatBarrier

\end{document}